\pdfoutput=1

\documentclass{llncs} 
\usepackage{url}
\usepackage{alltt}
\usepackage{amssymb}

\usepackage{epsfig}
\usepackage{wrapfig}
\usepackage{subfigure}

\usepackage{rotating}
\usepackage{moreverb}
\usepackage{fancyvrb}

\def\bd{\begin{description}}
\def\ed{\end{description}}
\def\bc{\begin{center}}
\def\ec{\end{center}}
\def\bq{\begin{quote}}
\def\eq{\end{quote}}
\def\bi{\begin{itemize}}
\def\ei{\end{itemize}}
\def\be{\begin{enumerate}}
\def\ee{\end{enumerate}}
\def\ba{\begin{array}}
\def\ea{\end{array}}

\newcommand{\chr}{{CHR}}

\newcommand{\lv}[1]{ }

\usepackage{stmaryrd}  
\newcommand{\parmapsto}{\mapsto}  

\newcommand{\myparagraph}[1]{\textbf{#1.}}

\begin{document}

\title{Exploring Parallel Execution Strategies for Constraint Handling Rules\\
{\normalsize Work-in-Progress Report}}

\author{%
Thom Fr{\"u}hwirth and Daniel Gall}

\institute{%
Ulm University, Ulm, Germany\\
\email{http://www.constraint-handling-rules.org}  
}

\date{\today}

\maketitle

\begin{abstract}
Constraint Handling Rules (CHR) is a declarative rule-based formalism and language. Concurrency is inherent as rules can be applied to subsets of constraints in parallel. Parallel implementations of CHR, be it in software, be it in hardware, use different execution strategies for parallel execution of CHR programs depending on the implementation language.

In this report, our goal is to analyze parallel execution of CHR programs from a more general conceptual perspective. We want to experimentally see what is possible when CHR programs are automatically parallelized. For this purpose, a sequential simulation of parallel CHR execution is used to systematically encode different parallel execution strategies. In exhaustive experiments on some typical examples from the literature, parallel and sequential execution can be compared to each other. The number of processors can be bounded or unbounded for a more theoretical analysis. As a result, some preliminary but indicative observations on the influence of the execution strategy can be made for the different problem classes and in general. 
\end{abstract}

\section{Introduction}

Constraint Handling Rules (CHR) 
\cite{fru_chr_book_2009,chr_survey_tplp10,chr-thesis-book} 
is both an effective concurrent declarative programming language and a versatile computational logic formalism.
In CHR, guarded reactive rules rewrite and augment a multiset of constraints. 
Concurrency is inherent, since rules can be applied to subsets of constraints in parallel.
The survey on parallelism and distribution in CHR \cite{fruehwirth2018parallelism}
reviews a variety of abstract and more refined semantics for parallel CHR as well as distributed CHR. 
These semantics come with several implementations in both software and hardware, many of them available online. 
These show promising experimental results with consistent parallel speed-up. 

\myparagraph{Minimum Example}
The following rule computes the minimum of some numbers 
given as multiset of constraints {\tt min($n_1$),min($n_2$),\ldots,min($n_k$)} (also referred to as \emph{state}):
\begin{verbatim}
min(N) \ min(M) <=> N=<M | true.
\end{verbatim}
The rule consists of a left-hand side, on which a pair of constraints has to be matched,
a guard check {\tt N=<M} that has to be satisfied, and an empty right-hand side denoted by {\tt true}.
In effect, the rule takes two {\tt min} candidates and removes the one with the
larger value (constraints after the \verb+\+ symbol are to be removed). 
Starting with a given initial state, CHR rules are applied exhaustively, resulting in a final state.
Note that CHR is a committed-choice language, i.e. there is no backtracking in the rule applications. 
Here the rule keeps on going until only one, the smallest
value remains as single {\tt min} constraint.

\myparagraph{Parallelism in CHR}
One of the main features of CHR is its inherent concurrency.
In a sequential computation, we apply one rule at a time.
Intuitively, in a parallel execution of a CHR program, rules can be applied simultaneously to
separate parts of a state. 
As we will see, CHR rules can even be applied in parallel to overlapping parts of a state,
in principle without the need to change the program. 
This is referred to as \index{logical parallelism}{\em logical parallelism} or 
\index{declarative concurrency}{\em declarative concurrency}.

In our {\tt min} example, the following computation is thus possible
(where we underline constraints involved
in a rule application):
\begin{center}
{\tt \underline{min(3)}, \underline{min(0)}, ~~~~~ \underline{min(2)}, \underline{min(1)} $\mapsto$}\\
{\tt \underline{min(0)}, ~~~~~~~~~~~~~ \underline{min(1)} $\mapsto$}\\
{\tt {min(0)}}\\
\end{center}
We arrive at the answer in less computation steps than with the sequential execution.
The rule can even be applied in parallel to
overlapping parts of the state, provided the overlap is not removed by any rule. 
For example, let the overlap be {\tt min(0)}.
Then the three pairs ({\tt min(0), min(3)}), ({\tt min(0), min(1)}) and ({\tt min(0), min(2)})
can be matched to different rule instances. 
They can be applied at the same time, since the common (overlapping) constraint {\tt min(0)} is not removed.
\begin{center}
{\tt \underline{\underline{\underline{min(0)}}},~~ \underline{min(3)}, \underline{min(2)}, \underline{min(1)} $\mapsto$}\\
{\tt {min(0)}}
\end{center}
So this is another, even shorter way to arrive at the same answer.

\myparagraph{Parallel Implementations}
The most popular parallel CHR implementations are in the lazy functional programming language Haskell
\cite{sulz_lam_parallelexecution_ppdp08,lam-concurrent-chapter2018} using its Software Transactional Memory (STM) extension.
CHR has also been implemented in specialized hardware \cite{triossi2012compiling} . 
The compiler translates the CHR code into the low-level hardware description language VHDL, 
which in turn creates the necessary hardware using Field Programmable Gate Array (FPGA) technology.
All implementations come with syntactic restrictions on rules. 
For example, most of them do not allow for propagation rules, i.e. rules that only add constraints.

In this paper, we present a sequential implementation of parallel CHR to automatically parallelize CHR programs. We then perform an experimental analysis on some well known programs from the literature to get a first impression of how CHR programs can be automatically executed in parallel. The sequential implementation allows us to easily exchange different parallel execution strategies, i.e. the strategies that decide which of the applicable rules are actually applied in a parallel transition step. Furthermore, it allows us to collect data about how many rules are applicable in theory and how many rules have actually been applied. 

In contrast to other parallel CHR implementations, our current experimental sequential implementation of parallel CHR supports all types of rules without any type of syntactic restrictions.
The implementations used fixed execution strategies inherited from the implementation language and hardware, respectively. Obviously, the number of processors is also bounded. In our experimental implementations, these limitations do not exist. This allows us to investigate what would be possible, if hardware was unlimited.
Even though we have produced more than 50 MB of data from test-runs, the number of problem instances and problem sizes documented in this work-in-progress report only allows for a preliminary, but we think nevertheless indicative interpretation of the experimental results observed. 

The paper is structured as follows: In Section~\ref{sec:chr-intro} we briefly introduce the sequential and parallel semantics of CHR. The sequential implementation of parallel CHR is then described in Section~\ref{sec:sequential-impl-par-chr}. Section~\ref{sec:exp-evaluation} discusses the experimental evaluation of the CHR example programs from the literature. 

\section{Parallel Abstract Operational Semantics of CHR}
\label{sec:chr-intro}

We will present the sequential equivalence-based abstract CHR semantics 
and extend it with parallelism. 
We just need a sequential transition that describes rule applications and another one that allows for parallel transitions. 
In this section, we assume basic familiarity with first-order predicate logic and state transition systems.
Readers familiar with CHR can skip most of this section.

\subsection{Abstract Syntax of CHR}

{\em Constraints} are relations, distinguished predicates of first-order predicate logic.
{\em Built-in (pre-defined) constraints} 
and
{\em user-defined (CHR) constraints} 
which are defined by the rules in a CHR program.
Built-in constraints can be used as tests in the guard as well as for auxiliary computations in the body of a rule.

{\rm
\myparagraph{Goals, States, Programs, Rules}
A {\em goal} is a conjunction of built-in and user-defined constraints.
A {\em state} is also a goal.
Conjunctions are understood as {\em multisets} of their conjuncts, i.e. in CHR states multiplicities of constraints play a role.
%
A {\em \chr\ program} consists of rules.  
A {\em (generalized) simpagation rule} has the form
\[r@ H_1 \backslash H_2 \Leftrightarrow C | B\]
where $r@$ is an optional {\em name} (a unique identifier) of a rule.
In the rule {\em head}, $H_1$ and $H_2$ are conjunctions of user-defined constraints,
the optional {\em guard} $C |$ is a conjunction of built-in constraints,
and the {\em body} $B$ is a goal.

In the rule, $H_1$ are called the {\em kept constraints}, while $H_2$ are called the {\em removed constraints}.
At least one of $H_1$ and $H_2$ must be non-empty. 
If $H_1$ is empty, the rule corresponds to a simplification rule, also written
$s@ H_2 \Leftrightarrow C | B.$
If $H_2$ is empty, the rule corresponds to a propagation rule, also written
$p@ H_1 \Rightarrow C | B.$
} 

\subsection{Abstract Operational Semantics of Parallel CHR}\label{sec:chr:semantics}

The semantics follows \cite{raiser_betz_fru_equivalence_revisited_chr09,betz2014unified}. 
It relies on a structural equivalence between states that abstracts away from technical details in a transition.

\myparagraph{Sequential Transition}
The equivalence relation $\equiv$ treats built-in constraints semantically and user-defined constraints syntactically.
Basically, two states are equivalent if they are logically equivalent (imply each other) 
while taking into account that user-defined constraints form a multiset. 
Using this state equivalence, the abstract CHR semantics is defined by a transition relation $\mapsto_r$ using only one single transition rule scheme. It defines the application of a rule $r$ in a CHR program $\mathcal{P}$ that rewrites a state $S$ to a state $T$. 
(Upper-case letters stand for (possibly empty) conjunctions of constraints in this section. The constraints in $G$ are the constraints in the states $S$ and $T$ that do not appear in the rule $r$. If the rule is clear from the context or does not play a role, we write $\mapsto$ instead of $\mapsto_r$.)
%
\begin{center}
{\bf (Apply)} \ \ \ $\underline{S \equiv (H_1 \land H_2 \land C \land G) \ \ \ \ (r @ H_1 \backslash H_2 \Leftrightarrow C | B) \in {\mathcal{P}}  \ \ \ \ \ \ \ \ (H_1 \land C \land B \land G) \equiv T}$\\
$S \mapsto_r T$
\end{center}
If the source state can be made equivalent to a state that contains the head constraints and the guard built-in constraints of a variant of a rule, then we delete the removed head constraints from the state and add the rule body constraints to it. 

Note that the abstract semantics does not account for termination of {propagation rules}.
In practice, propagation rules are not applied a second time to the same matching constraints.

\myparagraph{Parallel Transitions}
Following \cite{fru_parallel_union_find_iclp05}, CHR parallelism with overlaps can be defined as follows, 
see also Chapter 4 in \cite{fru_chr_book_2009}.
The transition {\bf (Parallel)} combines two transitions using conjunction into a single parallel transition 
where the overlap $E$ is kept. 
%
\begin{center}
%
{\bf (Parallel)} \ \ \ $\underline{A \land E \parmapsto C \land E \ \ \ \ \ \ B \land E \parmapsto D \land E}$\\
 \ \ \  \ \ \  \ \ \  \ \ \  \ \ \  \ \ \  \ \ \ $A \land B \land E \parmapsto C \land D \land E$\\
\end{center}
The correctness of the abstract parallel semantics can be established
by proving that
if $A \parmapsto B$ in parallel, then there exists a sequential computation $A \mapsto^* B$.
It suffices to show that the {\bf (Parallel)} transition can be simulated sequentially:
If    ${{A \land E}}$  $\mapsto$
  ${{B \land E}}$
and  ${{C \land E}}$  $\mapsto$
  ${{D \land E}}$,
then  ${{A} \land {C \land E}}$
  $\mapsto S \mapsto $ 
${{B} \land {D \land E}}$,
where $S$ is either $A\land D \land E$ or $B \land C \land E$.
We use this property of CHR parallelism to implement it faithfully in sequential CHR.

\section{Implementing Parallel CHR Sequentially}
\label{sec:sequential-impl-par-chr}

We implement the sequential simulation of parallel CHR with various execution strategies using an online CHR transformation tool
\url{http://pmx.informatik.uni-ulm.de/chr/translator/index.php}. The execution strategies decide the order in which applicable rules are applied in parallel.
The tool now includes a basic version of our parallel translation scheme described below. 
Originally, the straightforward tool was developed to allow for different sequential execution strategies (e.g. breadth-first, priorities, randomized) and to embed a variety of rule-based approaches in CHR (e.g. rewriting, functional programming, Petri nets, production rules).
It is used regularly in teaching at the University of Ulm. 
The rules below are somewhat edited to improve readability and succinctness.
In particular, system predicate calls for tracing and producing runtime statistics have been removed.

\myparagraph{Standard Program Transformation for Conflict Resolution}
Standard CHR applies rules as soon as they become applicable, without conflict resolution as can be found in classical production rules systems. In such systems, all applicable rules are collected first and a conflict resolution mechanism decides which of the applicable rules is chosen, e.g. by rule priorities or by random selection.
In order to embed other graph-based and rule-based approaches in CHR and to enable extensions of CHR by simple program transformation, conflict resolution for CHR rules has been implemented as follows (see also Chapter 6 in \cite{fru_chr_book_2009}).

Each original rule in the given program is split into two rules in the transformation to enable conflict resolution.
The first rule produces an instance of the applicable rule as a {\tt conflictset} constraint, without removing the head constraints of the matched rule.
The second rule actually applies the rule, provided the {\tt apply} constraint triggers this application. In the following code listing, this translation scheme is shown in more detail:
\begin{verbatim}
 % original rule
Keep \ Remove <=> Guard | Body.

 % standard program transformation for conflict resolution
Keep, Remove ==> Guard | 
       conflictset([rule(Strategy,Keep,Remove,Guard,Body)]).
Keep \ Remove, apply(rule(Keep,Remove,Guard,Body)) <=> Body.
\end{verbatim}
The first argument of {\tt rule}, which is {\tt Strategy}, specifies the execution strategy that should be used in conflict resolution.

The conflict resolution is implemented in essence by the following rules:
\begin{verbatim}
collect @ conflictset(L1), conflictset(L2) <=> 
          append(L2,L1,L3), conflictset(L3).
choose  @ fire, conflictset(L) <=> L=[_|_] | 
          choose(L,R,L1), conflictset(L1), apply(R), fire.
\end{verbatim}

The {\tt conflictset} constraints from the applicable rules are collected by rule {\tt collect} in a single conflict set as a list of rule instances.
If the trigger constraint {\tt fire} is present, the rule {\tt choose} will take the current non-empty conflict set and fire a rule from it.
The constraint {\tt choose} selects an applicable rule from the conflict set according to some strategy and returns also the remaining conflict set.
The rules are applied using {\tt apply}. This causes an extension of the conflict set with new rule instances. Finally, the removed constraint {\tt fire} is executed again to trigger another round of rule applications.
If the conflict set becomes empty, the computation terminates. Since in CHR implementations, propagation rules are only applied once to an identical combination of constraints, no rule is applied twice to the same store.

\lv{
\myparagraph{Parallel Execution Model}
Unlike previous parallel CHR implementations, we assume an unbounded number of processors in our simulation.
For each head matching, i.e. each combination of constraints from the store that matches the head, there is a processor handling the associated rule instance.
First, all processors check their rule guards in parallel to see if their rule is applicable.
Then, all processors with applicable rules proceed and try to apply their rules.
There may be conflicts because several processors may try to remove the same matched constraint.
Such conflicts can be handled with the usual techniques of e.g. locking or transactions, as has been done in the existing parallel CHR implementations.
In our sequential implementation this effect is achieved by globally ordering the rule applications, because we do not explicitly model processors.
With the conflict set, we explicitly represent potential rule applications instead.
}

\myparagraph{Rules for Parallel Execution Strategies}
For our purposes, the original program rules are slightly modified before the transformation.
The body constraints are wrapped into a delay constraint to prevent premature execution during the sequential simulation of a parallel execution step.
Furthermore, we extend the {\tt apply} constraint by one rule so that it can apply several rules instances, given in a list, in this way simulating one parallel computation step:
\begin{verbatim}
apply_all   @ apply([R|Rs]) <=> apply(R), apply(Rs).
\end{verbatim}
Note that rules may not be applicable, because a previous application may have removed the necessary constraints.

Below we show how the constraint {\tt choose} is implemented to simulate parallelism with different orderings among the applicable rule instances. The ordering corresponds to priorities among processors.
These strategies assume an unbounded number of processors.
\begin{verbatim}
par  @ choose(L,R,L1)  <=> L = [rule(par,_,_,_)|_]   | 
                       R=L, L1=[].
pars @ choose(L,R,L1)  <=> L = [rule(pars,_,_,_)|_]  | 
                       msort(L,R), L1=[].
pard @ choose(L,R,L1)  <=> L = [rule(pars,_,_,_)|_]  | 
                       msort(L,R1), reverse(R,R1), L1=[].
parr @ choose(L,R,L1)  <=> L = [rule(parr,_,_,_)|_]  | 
                       random_permutation(L,R), L1=[].
\end{verbatim}
In the guard, the strategy is determined by the first rule in the conflict set.
Strategy {\tt par} tries to execute applicable rules as they appear in the conflict set, without an extra ordering.
Strategy {\tt pars} and {\tt pard} sort the applicable rules first.
(Note that sorting can be done in constant time in parallel.)
Strategy {\tt parr} uses a random order.
Since all rule instances in the conflict set should be applied in parallel, {\tt choose} returns the empty list as a remainder.
If the number of processors is bounded instead, there are modified rules that try to apply just as many rules as there are processors and return the remaining conflict set in {\tt L1}.

\myparagraph{Complexity Considerations}
In our parallel execution model, the (unbounded) number of processors needed is equivalent to the number of head matchings.
Every constraint in the head of a rule can have multiple matching constraints in the store. The product of their multiplicities is an upper bound for the number of head matchings.
The actual number may be much smaller, depending on (functional) dependencies between constraints as enforced by common variables and guard conditions.
The number of applicable rule instances is further restricted by the guard of the rule.
The number of head matchings also provides an upper bound on the number of rules that can be applied in parallel.
If the rule removes constraints, then a typically much lower bound can be computed: it is the minimum of the number of matching removed head constraints.


\section{Experimental Evaluation of CHR Example Programs}
\label{sec:exp-evaluation}

Our exemplary CHR programs are mostly folklore in the CHR community, 
see e.g. Chapters 2 and 7 in \cite{fru_chr_book_2009}.
These are concise and effective 
implementations of classical algorithms and problems 
starting with finding primes, sorting, and
ending with Union-Find. 
They run in parallel without any need for modifying the program.
An exception is Union-Find, which is known to be hard to parallelize.
We can semi-automatically parallelize the algorithm with the help of CHR confluence analysis.

Most of these programs were already used in benchmarking parallel CHR implementations. 
In these implementations, a bounded number of processors linear to the number of constraints in the query was assumed. 
Under these assumptions, mostly optimal linear
parallel speed-ups were observed in a hardware implementation \cite{triossi2012compiling} and
in a software implementation \cite{sulz_lam_parallelexecution_ppdp08,lam-concurrent-chapter2018}.
In contrast, our work considers an unbounded number of processors and only simulates parallelism.
We do not consider absolute timings, but count the number of computation steps.
Through the program transformation approach our implementation is available with any Prolog CHR library.

Our experiments are parametrized by example, parallel scheduling strategy, number of processors and size of the problem.
Before a query is executed, its constraints are randomly permuted. This has an effect on the order of applicable rules in the conflict set.
Overall, we are interested in the number of computation steps, and in each of them, in the number of applicable rules and actually applied rules versus store size (number of constraints).
When the number of processors is bounded (including sequential execution with one processor), 
we ignore computation steps where none of the applicable rules considered is actually applied (because necessary constraints have been removed in the meantime). We consider such steps as garbage collection steps that can run in parallel with the main computation on a fixed limited number of processors. 

Our experiments were performed using the CHR implementation in SWI Prolog Version 7.4.2 under Ubuntu Linux 17.10.
So far, our experiments produced more than 50MB of test data, which is available at \url{http://www.uni-ulm.de/index.php?id=94438} for download.

\subsection{Algorithms of Erastothenes, Euclid, Fibonacci, von Neumann, Floyd and Warshall}

We describe some classical algorithms over numbers and graphs.
They are implemented as simple multiset transformations. 
Typically, they can be implemented with one kind of constraint and a single rule in CHR
that applies to pairs of constraints. These rules
can be applied in parallel to overlapping pairs of constraints.
Our running example of minimum falls into this category, as do the first following small programs.

\myparagraph{Prime Numbers}
The following rule is like the rule for minimum, but the guard is different, more strict.
In effect, it filters out multiples of numbers, similar to the Sieve of Erastothenes.
\begin{verbatim}
sift @ prime(I) \ prime(J) <=> J mod I =:= 0 | true.
\end{verbatim}
If all natural numbers from $2$ to $n$ are given as prime number candidates, only the primes within this
range remain, since non-prime numbers are multiples of other numbers greater or equal to $2$.

The number of head matchings is bounded by $n^2$.
According to a result of Dirichlet on the divisor problem, 
the sum of the number of divisors of numbers up to $n$ is approximated by $n log(n)$.
Thus the number of applicable rules after guard checking is about $n log(n)$. 
The number of applied rules is clearly less than $n$, otherwise no primes would be left.

In a parallel step, we can try to remove each number by associating it with another number such that the sift rule is applicable. 
It is therefore possible to compute all primes in a single parallel step.

In our experiments, given the  problem size parameter $n$, prime candidate numbers from $2$ to $n$ are given and randomly permuted. 
For an unbounded number of processors, the problem can be solved in one parallel step as predicted. When processors are bounded by the number of prime candidates $n$, two steps are required regardless of strategy. Thereby, 9.5 of 37 applicable rules are applied on average for bounded processors and 19 of 52 rules for unbounded processors ($n = 30$). The other applicable rules correspond to redundant removals of non-primes.
This means that limiting the number of processors to the problem size does not increase the complexity, it stays constant. Compare this to the observations for the minimum and greatest-common-divisor examples below.

\myparagraph{Minimum}
In this example from the introduction,
the minimum can be computed in one step when the number of processors are unbounded. 
The sequential execution has linear complexity (4/6/29 steps for $n = 5,7,30$). When processors are bounded by the initial number of minimum candidates $n$, only around half as many steps are needed (2/3/15 steps for $n = 5,7,30$). This can be explained by the fact that only on average around two rules can be applied in one parallel step when processors are bounded by $n$. 
So going from an unbounded to a bounded number of processors means to go from constant time to linear time complexity.

\myparagraph{Greatest Common Divisor}
The following rule computes the greatest common divisor (gcd) of natural numbers 
written each as {\tt gcd(N)}. 
\begin{verbatim}
gcd(N) \ gcd(M) <=> 0<N,N=<M | gcd(M-N).
\end{verbatim}
The rule replaces $M$ by the smaller number $M-N$ as in Euclid's algorithm.
The rule maintains the invariant that the numbers have the same greatest common divisor.
Eventually, if $N=M$, a zero is produced.
The remaining nonzero {\tt gcd} constraint contains the value of the gcd.

The number of head matchings is bounded by $n^2$.
Note that to any pair of non-zero gcd constraints, the rule will be applicable.
But since a constraint is removed in a rule application, less than $n$ rules can actually be applied
in a parallel computation step.
Maximally, all non-zero gcd constraint but one smallest gcd constraint will be removed.
In the worst case, the parallel computation deteriorates to a sequential one. This is the case if only two non-zero gcd constraints are left, one containing a large number and the other a very small number.
Overall, this means that the number of parallel steps is bounded by the largest number occurring in the problem.
Parallel execution may lead to smaller similar numbers more quickly than a sequential implementation. 
Indeed, a super-linear speed-up was observed in the parallel software implementation of CHR in \cite{lam-concurrent-chapter2018}.

In our experiments for gcd, we either use a range of numbers from $2$ to $n$ (examples \emph{gcd}) or rounded numbers derived 
from $1.618^k+k$, for $k$ in this range (examples \emph{gcd2}). For bounded processors, approximately 2 rules can be applied in a parallel step ($n = 7$ (\emph{gcd} and \emph{gcd2}) and $n = 30$ (\emph{gcd})). The problem generates a quadratic number of applicable rules (mean between 800 and 1200 for $n = 30$) as predicted. For unbounded processors, between 9 and 15 rules ($n = 30$, \emph{gcd}) or 2 and 5 rules ($n = 7$, \emph{gcd}) could be applied per step. It is noticeable that the random runs needed the least number of steps for \emph{gcd} with $n = 30$. 
This indicates that the order can play an important role and confirms the super-linear speed-up observed in other parallel CHR implementations.
Limiting the number of processors seems to introduce a penalty factor of $n$ as for minimum.

\myparagraph{Fibonacci Numbers}
Next we show a naive but highly parallel CHR program that computes the value of the $n$-th Fibonacci number. 
\begin{verbatim}
0   @ findFibo(0) <=> sum(1).
1   @ findFibo(1) <=> sum(1).
n   @ findFibo(N) <=> N>1 | findFibo(N-1), findFibo(N-2).          
sum @ sum(N1), sum(N2) <=> sum(N1+N2). 
\end{verbatim}
In $n$ parallel steps, the first three rules for {\tt findFibo} will produce an exponential number of sum constraints.
This leads to a double exponential number of head matchings for sum constraints.
It is actually less, because the computations of findFibo and sum happen in parallel to some extent.
In each parallel step, the number of sum constraints can be halved.
Therefore the number of additional steps for summing is bounded by $n$ as well.

In our experiments, $n$ is the parameter for a single findFibo constraint in the query. 
Due to the double exponential number of applicable rules, our experiments were limited to small problem sizes. This makes it hard to verify our predications.
The experiments indicate that on average only between 1.8 and 2.5 rules can be applied per step for bounded processors ($n = 5,7$, maximum applied rules were 5 and 7, respectively, over all problem instances). For unbounded processors, on average 2.75/5.55 ($n = 5,7$) rules can be applied per step. The strategies do not have a significant influence on the number of steps. 

\myparagraph{Merge Sort}
Next we present a variation of von Neumann's merge sort algorithm for a set of numbers.
The initial goal state contains arcs of the form {\tt v->V} for each value {\tt V},
where {\tt v} is a given smallest (dummy) value.
\begin{verbatim}
msort @ A->B \ A->C <=> A<B, B<C | B->C.
\end{verbatim}
The rule updates the first argument of the second arc constraint matched.
It is replaced by a larger value and the two resulting arcs form a small chain
{\tt A->B,  B->C}.
The rule maintains the invariant for each arc that its first argument is smaller than its second argument.
Eventually, in each arc, a number will be followed by its immediate successor, 
and thus the resulting chain of arcs is sorted.

The sequential complexity of the program is $n^2$, since in each rule application, only the first argument changes. It becomes larger, but never as large as the second argument. There are $n$ constraints with at most $n$ different values, which leads to the quadratic complexity. 
(Optimal $nlog(n)$ complexity can be achieved by introducing a second rule that restricts merging to arc chains of roughly the same size.)
In a parallel step,
there are at most $n^2$ head matchings (about half of them applicable) and at most $n$ rules can actually be applied.
So in the best case, we just need up to $n$ steps.

\begin{figure}
  \centering
  \subfigure[Processors unbounded ($n = 30$)]{\includegraphics[width=0.49\linewidth]{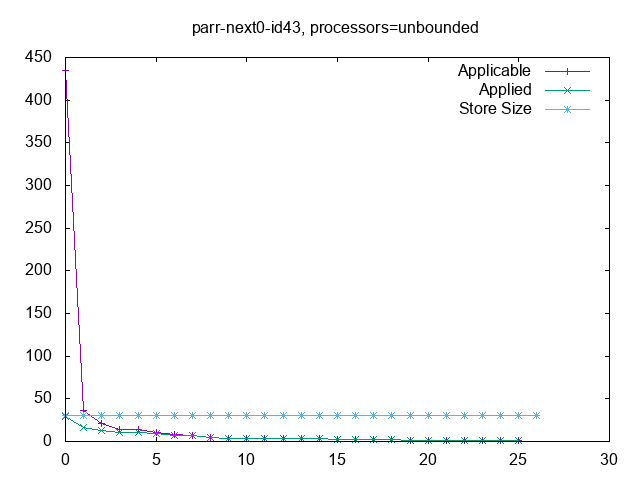}}
  \subfigure[Processors bounded by 30 ($n = 30$)]{\includegraphics[width=0.49\linewidth]{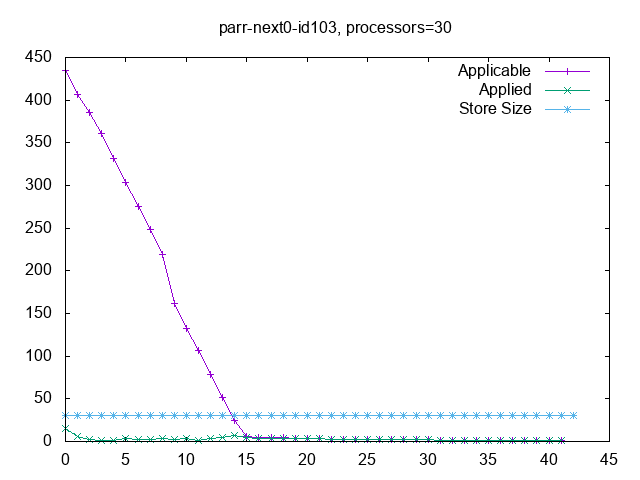}}
  \caption{Parallel execution of Merge Sort with random strategy. For each transition step on the $x$-axis the number of applicable rules, actually applied rules and the store size is plotted on the $y$-axis.} 
  \label{fig:exp:merge}
\end{figure}

In our experiments, we use a range of numbers from $2$ to $n$. (Recall that the numbers are randomly permuted in the query.) 
For unbounded processors, the algorithm takes around $n$ transition steps (between 26 and 29 steps for $n = 30$) 
which is optimal with regard to our theoretical considerations. 
The strategy does not seem to play an important role.
%
For bounded processors, between 40 and 48 steps are needed for $n = 30$.  
An example run for $n = 30$ can be seen in Figure~\ref{fig:exp:merge}.
Clearly more experiments with different problem sizes are needed.

\myparagraph{Floyd-Warshall All-Pair Shortest Paths}
The rules find the shortest path distance between connected pairs of nodes in
a directed graph whose edges are annotated with non-negative distances.
\begin{verbatim}
elim  @ path(X,Y,D1) \ path(X,Y,D2) <=> D1<D2 | true.

base  @ arc(X,Y,D) ==> path(X,Y,D).
trans @ arc(X,Y,D1), path(Y,Z,D2) ==> X\=Z | path(X,Z,D1+D2).
\end{verbatim}
The rule {\tt elim} removes the larger path between two nodes.
In the sequential execution it is typically assumed that for a path,
the rule {\tt elim} is always tried before the rule {\tt trans}.
In the parallel execution this cannot be guaranteed, and therefore more rule applications are possible.

If $d$ is the number of different distances, the rule {\tt elim} can be applied at most $d$ times to the same pair of nodes {\tt X,Y}.
Rule {\tt base} creates a path for each arc.
With $n$ nodes, $a$ arcs and $d$ distances, rule {\tt trans} can be applied up to $nad$ times, dominating the complexity.

In our experiments, we randomly generate graphs consisting of $n$ nodes and $2n$ or $3n$ arcs with ordered random pairs of nodes. 
To make the problem nontrivial, each pair is associated with a unique distance that increases quadratically with the difference in their node values. 
This means that longer paths can have shorter distances than paths of less length and direct arcs.
Since longer paths can only be computed later, this will cause a chain of recomputations triggered by this longer path with shorter distance.

The experiments indicate that for unbounded processors the strategy does not play an important role for the number of transition steps: The algorithm always used a linear number, 6 or 7 steps, for all randomly generated instances and all strategies ($n = 7$). For bounded processors, the number of steps increases considerably. Interestingly, the sorting strategies seem to be effective: They yield 36 steps for \emph{pars} and 25 for \emph{parsd} (note that those are also different instances) compared to 78/119/61/80 for the other strategies (again run on different instances). 
We do not yet have an explanation for these observations. 
Obviously the problem instances do not allow for worst-case behavior in this case.
We repeated the experiment for the outlier with 119 steps and a random strategy. This instance only needed 81 and 83 steps with pars and parsd respectively (72/96/106/95 steps for the other strategies\footnote{The re-execution of the experiment with the same problem instance has been performed on permuted start states. The \emph{parr} runs used different random numbers.}).
More experiments with different problem instances and sizes are needed.


\myparagraph{SAT Solving}
The SAT formula is given as a tree of its sub-expressions. The tree nodes are of the form {\tt eq(Id,B)},
where {\tt Id} is a node identifier and {\tt B} is either a Boolean variable written {\tt v(X)}
or a Boolean operation ({\tt neg, and, or}) applied to identifiers. 
Additionally, a {\tt f(L,[])}
constraint is required in the initial state,
where {\tt L} is a list of all $n$ variables in the SAT formula. 
{\small
\begin{verbatim}
generate @ f([X|Xs], A) <=> f(Xs,[true(X)|A]), f(Xs,[false(X)|A]).

assign @ f([],A), eq(T,v(X)) ==> member(true(X),A)  | sat(T,A,true).
assign @ f([],A), eq(T,v(X)) ==> member(false(X),A) | sat(T,A,false).

sat(T1,A,S1), eq(T,neg(T1)) ==> neg(S1,S), sat(T,A,S).
sat(T1,A,S1), sat(T2,A,S2), eq(T,and(T1,T2)) ==> and(S1,S2,S), sat(T,A,S).
sat(T1,A,S1), sat(T2,A,S2), eq(T,or(T1,T2))  ==> or(S1,S2,S), sat(T,A,S).
\end{verbatim}
}
The {\tt generate} rule generates, in $n$ parallel steps, $2^n$ {\tt f} constraints
representing all possible truth assignments to variables as a list in its second argument. 
Then, in one more parallel step using the {\tt assign} rules, all 
variables in the given formula are assigned truth values for each
assignment, represented by {\tt sat} constraints. 
Finally, the remaining three rules
determine the truth values of all sub-expressions of the formula bottom-up. 
Therefore, the number of parallel steps in this phase is bound by the depth of the formula. 

In our experiments, we use $n$ variables and generate a random formula of depth $2$
with $n$ boolean operations on these variables, with 
about half of them disjunction and conjunction and one negation on average.

\begin{figure}[htb]
  \centering
  \subfigure[Processors bounded by 5 ($n = 5$)]{\includegraphics[width=0.49\linewidth]{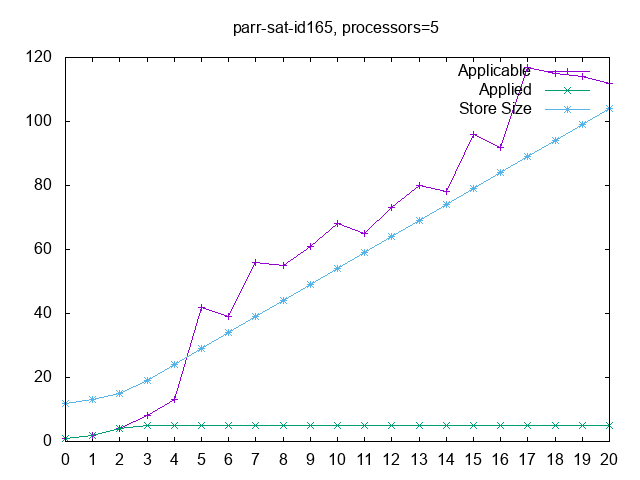}}
  \subfigure[Processors bounded by 7 ($n = 7$)]{\includegraphics[width=0.49\linewidth]{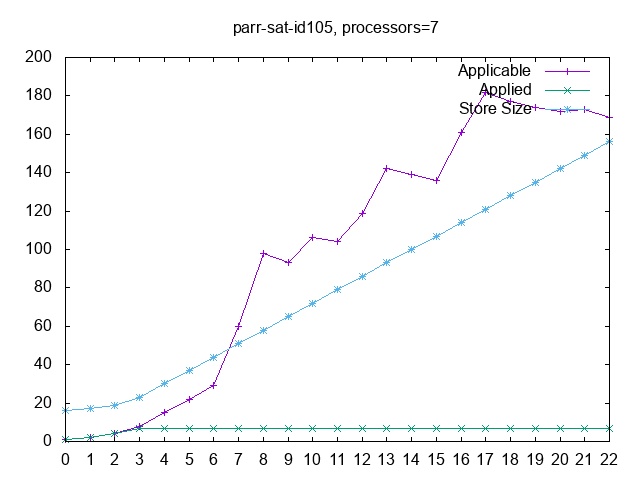}}
  \subfigure[Processors unbounded ($n = 5$)]{\includegraphics[width=0.49\linewidth]{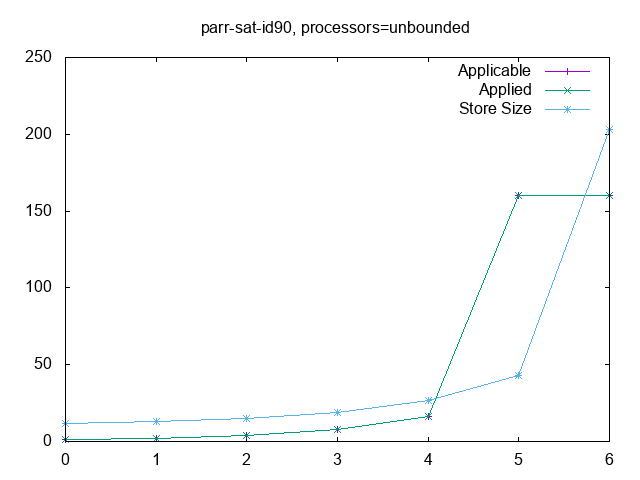}}
  \subfigure[Processors unbounded ($n = 7$)]{\includegraphics[width=0.49\linewidth]{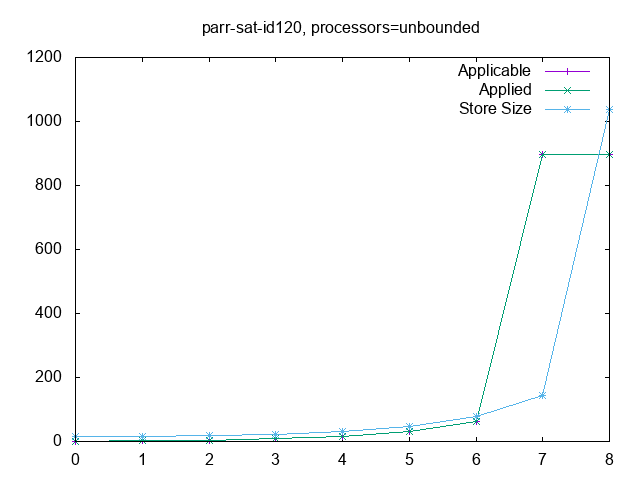}}
  \caption{Parallel execution of the SAT solver with random strategy. For each transition step on the $x$-axis the number of applicable rules, actually applied rules and the store size is plotted on the $y$-axis.} 
  \label{fig:exp:sat}
\end{figure}

Due to the exponential number of applicable rules, our experiments were limited to small problem sizes. 
%
For unbounded processors, we have observed that the applicable rules are applied exhaustively in our experiments, i.e. that all applicable rules have been applied in each step. For bounded processors, the program starts with applying one rule and increases the number of applied rules per step until the maximum number of processors is reached. This can be seen in Figure~\ref{fig:exp:sat} for four instances with $n = 5,7$ and bounded and unbounded processors. 
Note the different scales and that applied and applicable rules coincide in the unbounded case.

\subsection{Classical Algorithms with Statefulness}

These algorithms about abstract problems are characterized by their {\em statefulness}, i.e. their essence is a state change, an update.
While other purely declarative languages may not have an efficient way to update, 
CHR has a proven one by constant-time updating (i.e. removing and adding) user-defined constraints
\cite{sney_schr_demoen_chr_complexity_toplas09}.
Of course, arbitrary updates pose a problem for parallelisation because they may be in conflict.
Indeed, the problem is NP-complete in the presence of deadlocks.

\myparagraph{Blocks World}
Blocks World is a classical planning problem in Artificial Intelligence. It simulates robot arms re-arranging stacks of blocks.
{\small
\begin{verbatim}
grab  @ grab(R,X),  empty(R), clear(X), on(X,Y) <=> hold(R,X), clear(Y).
putOn @ putOn(R,Y), hold(R,X), clear(Y) <=> empty(R), clear(X), on(X,Y).
\end{verbatim}
}
The {\em operation constraints} {\tt grab} and {\tt putOn} specify the action that is taken.
The other constraints are {\em data constraints} holding information about the scenario.
Operation constraints update data constraints.
The rule {\tt grab} specifies that robot arm {\tt R} grabs block {\tt X} if 
{\tt R} is empty and block {\tt X} is clear on top and on block {\tt Y}.
As a result, robot arm {\tt R} holds block {\tt X} and block {\tt Y} is clear.
The rule {\tt putOn} specifies the inverse action.
The {data constraints} in the rule switch sides.
At any time, only one of the actions is thus possible for a given robot arm.
Parallelism is induced by introducing several robot arms. 

In our experiments, we initially have $n$ empty robot arms and $n$ or $2n$ blocks on stacks of size $2$.
Each robot arm grabs a different block.
Some blocks may not be immediately available, because another block is on top of them.
Finally, there are $n$ random put constraints. 
This means that there may be conflicting put constraints for the same arm or for the same target block.
In our problem instances, we expect that most blocks can be grabbed and then put down simultaneously.
Worst-case behavior could be achieved when the same block has to be grabbed by all robot arms. 
That would enforce a sequential computation.

\begin{figure}
  \centering
  \subfigure[Processors bounded by 30 ($n = 30$)]{\includegraphics[width=0.49\linewidth]{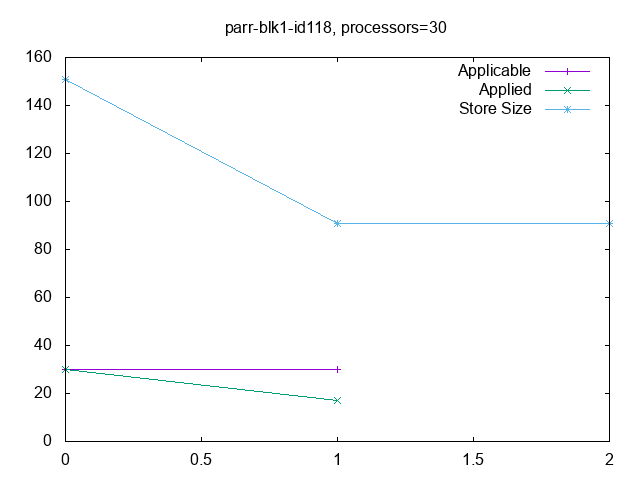}}
  \subfigure[Processors unbounded ($n = 30$)]{\includegraphics[width=0.49\linewidth]{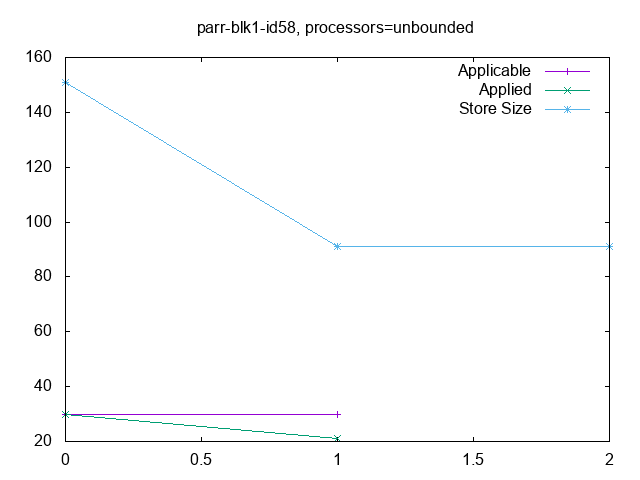}}
  \caption{Parallel execution of the Blocks World example with random strategy. For each transition step on the $x$-axis the number of applicable rules, actually applied rules and the store size is plotted on the $y$-axis.} 
  \label{fig:exp:blocks}
\end{figure}

The experiments revealed that the number of applied rules is very close to the number of applicable rules. The strategy does not have a high impact and the runs with bounded processors are comparable to the runs with unbounded processors. This is the case because the problem is bound to maximally $n$ available robot arms. This means that there are at most $n$ applicable rules and hence at most $n$ processors can be used. In Figure~\ref{fig:exp:blocks} a bounded and an unbounded run are shown. Furthermore, the experiments suggest that the random execution strategy \emph{parr} may be most efficient. This should be analyzed for more instances.

\myparagraph{Parallel Union-Find Algorithm} 
Tarjan's classical union-find (UF) algorithm efficiently
maintains disjoint sets under the operation of union. 
Each set is represented by a rooted tree, whose
nodes are the elements of the set.
In \cite{fru_parallel_union_find_iclp05}, we implement the UF algorithm in CHR with optimal 
time and space complexity and with the anytime online algorithm properties of CHR. 
For space reasons, we only discuss the basic Union-Find (UF) algorithm here.
In CHR, the {\em data constraints} \texttt{root} and arc \texttt{->}
represent the tree data structure.
With the UF algorithm come several {\em operation constraints}:
{\tt find} returns the root of the tree in which a node is contained,
{\tt union} joins the trees of two nodes,
{\tt link} performs the actual join.
{\small
\begin{verbatim}
union    @ union(A,B) <=> find(A,X), find(B,Y), link(X,Y).

findNode @ A->B  \ find(A,X) <=> find(B,X).
findRoot @ root(A) \ find(A,X) <=> found(A,X). 

linkEq   @ link(X,Y),found(A,X),found(A,Y) <=> true.  
linkRoot @ link(X,Y),found(A,X),found(B,Y),root(A) \ root(B) <=> B->A.
\end{verbatim}
}
The second argument of the find operation \texttt{find} holds a fresh variable as identifier.
When the root is found, it is recorded in the constraint \texttt{found}.

Union-Find is hard to parallelize.
Indeed, CHR {confluence analysis} 
reveals a {deadlock}:
when we are about to apply the {\tt linkRoot} rule,
another link operation may remove one of the roots that we need. 
From the non-confluent states we can derive an additional rule for {\tt found} (it mimics the rule 
{\tt findNode}):
{\small
\begin{verbatim}
foundUpdate @ A->B \ found(A,X) <=> found(B,X).
\end{verbatim}
}
%
While obvious here, this confluence-based approach 
yields a non-trivial parallel variant of the optimized UF algorithm with path compression \cite{fru_parallel_union_find_iclp05}.
 
In the algorithm, each union leads to two find operations. If the arcs form a chain, they need linear time in the worst case to reach the root node.
The introduction of {\tt found} does not change this complexity.
In the basic UF algorithm, the worst case complexity for $n$ nodes and $u$ unions is therefore $nu$. 

In our experiments, we have $2n$ root nodes with either $2n$ random unions between these nodes
or $n$ random unions where each node occurs exactly once. 
%
We observed in the experiments that the number of applied rules per step was very close to the applicable rules for unbounded processors and was in the same dimension even for bounded processors. This can also be observed in the number of transition steps for unbounded and bounded processors that is almost identical over all problem instances and strategies. This indicates that the CHR implementation of the algorithm can be parallelized successfully. 

\section{Conclusions and Future Work}
\label{sec:conclusion}

In our work-in-progress report, we presented first experimental results on comparing parallel execution strategies for CHR using a straightforward sequential implementation to simulate parallelism. 
Unlike existing implementations of parallel CHR, our implementation allows for an unbounded number of processors.
It supports all types of CHR rules without any type of syntactic restrictions. Our approach is the first step to see what is possible when CHR programs are automatically parallelized.

We implemented the sequential simulation of parallel CHR supporting various execution strategies 
using an online CHR source-to-source transformation tool that adds conflict resolution to CHR.
Through the program transformation approach our implementation is available with any Prolog CHR library.
A basic version of our translation scheme is available at the online tool \url{http://pmx.informatik.uni-ulm.de/chr/translator/index.php}. 

Our experiments are parametrized by example, parallel execution strategy, number of processors and size of the problem.
Most of our example programs were already used in benchmarking parallel CHR implementations.
We do not consider absolute timings, but count the number of computation steps.
The different execution strategies reorder the constraint in the problem. This corresponds to distribute the applicable rules among the given processors that have different priorities for removing constraints due to rule applications.
 
Overall, straightforward parallel execution of existing CHR programs often leads to optimal linear speed-ups in the sense of reduced computation steps.
This was already observed for the existing parallel implementations of CHR in the literature.
Simple problems like finding the minimum or prime numbers can in principle run in constant time given enough processors.
Some algorithms use the available processors (almost) exhaustively (e.g., SAT, Blocks World, Union-Find).
Especially the results of the Union-Find algorithm are promising.
When the number of processors is bounded, there are also algorithms where only a speed-up of factor 2 over a sequential execution can be observed
(e.g. Prime Sieve, Fibonacci, GCD). 

Although we only integrated generic execution strategies like sorting for processor priorities or randomization of the order of the constraints in the problem, our experiments indicate that the resulting order of rule applications can play an important role for effectiveness
(e.g., GCD, Blocks World, Floyd-Warshall). 

Even though we have produced more than 50 MB of data from test-runs, 
it allows only for a preliminary interpretation of the experimental results.
The test data is available at \url{http://www.uni-ulm.de/index.php?id=94438} for download.
To investigate these issues further, it would be important to generate more problem instances of different structure and size, and compare the different strategies with equivalent start states. Furthermore, some more problem-specific execution strategies may be tested. We plan to investigate if there exists a meta-complexity theorem for parallel CHR execution similar to the one in the sequential case \cite{fru_complexity2_entcs02}.
%

\bibliographystyle{abbrv}           
\bibliography{par-chr-sim,CHR2015,chr-biblio-jan-2017}

\label{lastpage}

\end{document}